\documentclass{aa520}
\usepackage{txfonts}
\usepackage{graphicx}
\usepackage[latin1]{inputenc}
\def\HII{H\,{\sc{ii}}}
\def\SII{S\,{\sc{ii}}}
\def\HI{H\,{\sc{i}}}
\def\fs{\hbox{$.\!\!^{\rm s}$}}

\def\farcm{\hbox{$.\mkern-4mu^\prime$}}
\def\farcs{\hbox{$.\!\!^{\prime\prime}$}}
\def\arcmin{\hbox{$^\prime$}}
\def\arcsec{\hbox{$^{\prime\prime}$}}

\def\degr{\hbox{$^\circ$}}
\def\h{\hbox{$^{\reset@font\r@mn{h}}$}}
\def\m{\hbox{$^{\reset@font\r@mn{m}}$}}
\def\s{\hbox{$^{\reset@font\r@mn{s}}$}}

\begin{document}

   \title{Sequential star formation at the periphery of the 
   H\,{\large II} regions Sh~217 and Sh~219\thanks{Based on 
   observations done at the 
   Observatorio Astronómico National at San Pedro Mártir, México, and 
   at the Observatoire de Haute-Provence, France.}
   \thanks{Tables 3 and 4 are available only at the CDS, via 
   anonymous ftp to cdsarc.u-strasbg.fr (130.79.128.5)
   or via http://cdsweb.u-strasbg.fr/cgi-bin/qcat?J/A+A/(vol)/(page)}
   }
   \titlerunning{Sequential star formation}
   \subtitle{}
   \author{L.~Deharveng\inst{1}
          \and
          A.~Zavagno\inst{1}
          \and
          L.~Salas\inst{2}
          \and
          A.~Porras\inst{3}
          \and
          J.~Caplan\inst{1}
          \and
          I.~Cruz-González\inst{4}
          }
   \authorrunning{L.~Deharveng et al.}
   \offprints{L.~Deharveng}

   \institute{
      Laboratoire d'Astrophysique de Marseille, 2 Place Le Verrier, 13248 Marseille
      Cedex 4, France\\ lise.deharveng@oamp.fr\\
      \and
      Observatorio Astronómico Nacional, IA-UNAM, 
      Apdo.\ Postal 877, Ensenada, B.C.\ México\\ 
      salas@astrosen.unam.mx\\
      \and
      Instituto de Astronomía, UNAM, Apdo.\ Postal 3--72 (Xangari), 58089 Morelia,
      Mich.\ México\\
      a.porras@astrosmo.unam.mx\\
      \and
      Instituto de Astronomía, UNAM, Apdo.\ Postal 70-264, Cd. Universitaria,
      04510 México, D.F. México\\ 
      irene@astroscu.unam.mx\\  
      }
   
   \date{Received; accepted}
   
     \abstract{     
     The \HII\ regions Sh~217 and Sh~219 are textbook examples of a
     Strömgren sphere surrounded by an annular photodissociation region 
     (PDR). The annular PDR is observed in both the 21~cm atomic hydrogen
     emission and the dust (PAH) emission near 8~$\mu$m (MSX Survey).

     An ultracompact radio continuum source is observed in the direction of
     the annular PDR, in both Sh~217 and Sh~219. $JHK$ observations
     show the presence of highly reddened stellar clusters 
     ($A_V \sim 20$ mag) in the directions of these radio sources. These
     clusters are also IRAS sources, of luminosities 22\,700~$L_\odot$
     for Sh~217 and 5900~$L_\odot$ for Sh~219. Each cluster contains at
     least one luminous star with an IR colour excess; the one in the
     Sh~219 cluster shows H$\alpha$ emission. The cluster 
     associated with Sh~217 is almost
     spherical and contains luminous objects at its centre. The cluster
     associated with Sh~219 is elongated along the ionization front of
     this \HII\ region.

     We argue that these are `second-generation clusters', which means
     that the physical conditions present in the PDRs, 
     close to the ionization fronts, have favoured the 
     formation of clusters containing massive objects. We discuss the 
     physical mechanisms which may be at the origin of the observed 
     triggered star formation.

	}
\maketitle

\section{Introduction}

Most massive stars are observed associated with dense stellar clusters
(Testi et al.\ 1999, 2001 and references therein). One possible
interpretation of this observational fact is that the mechanism(s) for
massive-star formation requires a cluster potential; the stars may form by
coalescence from intermediate mass protostars (Bonnell et al.~1998,
2001, Stahler et al.~2000) or by accretion (Bernasconi \& Maeder 1996,
Norberg \& Maeder 2000, Behrend \& Maeder 2001), at the centres of
dense clusters. However, the observational data do not require such an
interpretation; they are also compatible with the statistics of
randomly assembled stars, with masses distributed according to the
standard Initial Mass Function (Bonnell \& Clarke 1999). Thus the mode
of formation of massive stars is still debated. It is an important
observational challenge to identify the progenitors of such stars and
to characterize the physical conditions required to form them
or the protoclusters.

Dobashi et al.~(2001) have compiled a list of about 500 molecular
clouds harbouring IRAS point sources with colours typical of
protostars. Their study shows that protostars in clouds associated with
\HII\ regions are more luminous than those in clouds far from \HII\
regions, thus highlighting the part played by \HII\ regions in the
formation of massive objects (stars or clusters). Observational
evidence that \HII\ regions are efficient sites of {\it 
second-generation} star formation is presented by Elmegreen (1998). 
Several physical mechanisms, discussed by Whitworth et al.\ (1994) 
and Elmegreen (1998), have been proposed to account for triggered star
formation at the periphery of \HII\ regions. These mechanisms depend on
the high pressure in the ionized gas and on the expansion of the \HII\
region: i) Star formation may result from the radiation-driven
implosion of a pre-existing molecular clump: cometary globules are
formed, surrounded by optically bright rims (Bertoldi 1989, Lefloch \&
Lazareff 1994). Indeed, star formation is observed at the centre of these
globules (Lefloch et al.~1997, Sugitani et al.~2000 and references
therein, Ogura et al.~2002). ii) Or, as described in the model of
Elmegreen and Lada (1977, 1978; see also Whitworth et al.\ 1994), a
compressed layer of high-density neutral material forms between the
ionization front and the shock front preceding it in the neutral gas,
and star formation occurs when this layer becomes gravitationally
unstable.

We present here the case of the two \HII\ regions Sh~217 and Sh~219,
with their nearby associated clusters. Their simple morphology 
may help us to understand the process of sequential star
formation possibly at work here. Each region appears to be a Strömgren
sphere centred on a single exciting star, and surrounded by an annular
photodissociation region (PDR). In each case a young cluster,
containing at least one massive star, is observed at one position on
the PDR annulus, like a brilliant on a ring. In Sect.~2 we describe
the two \HII\ regions, what is known about their exciting stars and the
morphology of the photoionized and photodissociated zones. In Sect.~3
we present our near-IR observations of these newly-discovered young
clusters. In Sect.~4 we discuss the process of sequential star
formation in the light of what is observed in Sh~217 and Sh~219.
Conclusions are drawn in Sect.~5.

\section{Description of the \HII\ regions}

Sh~217 and Sh~219 are optically visible \HII\ regions (Sharpless
1959) separated by some $45\arcmin$. The J2000 coordinates of their
central exciting stars are, for Sh~217,
$\alpha=4^{\rm h}58^{\rm m}45\fs4$, 
$\delta=+47{\degr}59{\arcmin}55{\arcsec}$
and, for Sh~219, 
$\alpha =4^{\rm h}56^{\rm m}10\fs5$, 
$\delta=+47{\degr}23{\arcmin}34{\arcsec}$.

\subsection{The distances}

Although Sh~217 and Sh~219 are seen in projection at the edge of the
supernova remnant HB9 (G160.9+2.6), these objects are not related to
HB9, which is at a distance of about 1.2~kpc (Leahy \& Roger 1991) 
whereas Sh~217 and Sh~219 have distances between 4 and 6~kpc. As we
shall see, it is not clear whether Sh~217 and Sh~219 lie at the 
same distance.

In the following we discuss the photometric distances of their
exciting stars.

The main exciting star of Sh~219 is of spectral type B0V, with
$V$=12.10 and $B-V$=0.53 (Georgelin et al.\ 1973, Moffat et al.\
1979, Lahulla 1987). Using the colour calibration of Schmidt-Kaler 
(1982) we estimate a visual extinction of 2.57 mag. The calibrations
of Vacca et al.\ (1996) and of Humphreys \& McElroy (1984) for the
absolute magnitude of a B0V star lead to distance estimates of 
5.6~kpc and 4.6~kpc, respectively. (The absolute magnitude is
particularly uncertain for this spectral type.)

The spectral type of the main exciting star of Sh~217 is uncertain --
O9.5V, B0V or O8V according to Georgelin et al.\ (1973), Moffat et
al.\ (1979) and Chini \& Wink (1984) respectively; consequently its
distance is very uncertain. In Sect.~2.3 we explain why we favour
an O9V or later spectral type. Assuming O9V and using the absolute
magnitude calibrations of Vacca et al.\ (1996) or of Schmidt-Kaler 
(1982) (which are not very different), its observed magnitude
$V$=11.34 and colour $B-V$=0.40 correspond to a visual extinction of
2.20 mag and a distance in the range 5.18--5.35~kpc.

Thus the distance of Sh~219 is uncertain because of the uncertainty
in the absolute magnitude of a B0V star, and the distance of Sh~217
is uncertain because of the uncertainty in the spectral type of its
exciting star.

The kinematic distances deduced from the velocities of the associated
molecular material (Table~1) by using the Galactic rotation curve of
Brandt \& Blitz (1993) are 5.2~kpc for Sh~217 and 4.2~kpc for 
Sh~219.

Given all these uncertainties, we adopt, in the following discussion,
a distance of 5.0$\pm$0.8~kpc for both Sh~217 and Sh~219.

\begin{table}
\caption{Kinematic information on Sh~217 and Sh~219}
\begin{tabular}{cccl}
 \hline\hline
  \hspace*{1mm}  & \multicolumn{2}{c}{$V_{\rm LSR}$ (km s$^{-1}$)} & \hspace*{1mm} \\
        & Sh~217 & Sh~219 & references  \\
  \hline
  CO       & $-$20.5       & $-$24.5      & Blitz et al.\ 1982 \\
  CO       & $-$18.1       & $-$25.0      & Wouterloot \& Brand 1989 \\        
  CO       & $-$18.2       &              & Jackson \& Sewall 1982 \\   
  \HI\     & $-$17         & $-$23        & Roger \& Leahy 1993 \\
 H$\alpha$    & $-$20.4        & $-$31.0      & Fich et al.\ 1990 \\
 \hline
\end{tabular}
\end{table}

\subsection{The morphology of the H\,{\small II} regions}

Sh~219, in Fig.~1a, is a prototypical Strömgren sphere: in the
optical it appears to be a nearly spherical \HII\ region centred on its
exciting star. In the radio continuum, shown in Fig.~1b, it also
appears almost circular, of diameter 4.4~pc ($3 \arcmin$) if situated at 5~kpc (Fich
1993 and references therein; Leahy 1997). The 1.4~GHz radio continuum 
(ionized gas) map from the NVSS survey (Condon et al.\ 1998) shows a
slight asymmetry~-- an extension of the emission towards the 
south-west. Leahy (1997) has detected a non-resolved radio source in 
this direction, as shown in Fig.~1b; the J2000 coordinates of this
source (estimated from Leahy's map) are:
$\alpha =4^{\rm h}56^{\rm m}02\,\fs2 \pm 0\,\fs4 $,
$\delta=+47{\degr}23{\arcmin}06{\arcsec} \pm 4\arcsec$. Sh~219's
electron density estimated from the intensity ratio of the
[O\,{\sc{ii}}] 3726--3727~\AA\ doublet is rather low -- $\sim$170~cm
$^{-3}$ (Deharveng et al. 2000) -- and its rms electron density
estimated from the radio continuum emission is 55~cm$^{-3}$ (Roger \&
Leahy 1993).

However, a circular appearance does not prove spherical symmetry; a
champagne-model \HII\ region (Tenorio-Tagle 1979) viewed face-on
would appear symmetric for an observer situated on its axis. In the
case of Sh~219, the champagne model is suggested by comparison of the
radial velocities obtained for the ionized gas and for the associated
molecular material. Table~1 shows that the ionized gas flows away
from the molecular cloud, more or less in the direction of the
observer, with a velocity of $\sim$6.5~km~s$^{-1}$, in agreement with
the predictions of the champagne model.

Sh~217, in Fig.~2a, is somewhat elliptical, with outer dimensions
of 10~pc $\times$ 14~pc. Its south-west part is the brightest, both
at optical and radio wavelengths. In the radio continuum, in 
Fig.~2b, it appears as a half-shell surrounding the exciting star 
(Fich 1993, Roger \& Leahy 1993, Condon et al.\ 1998). Here again an
ultracompact radio source lies at its south-west periphery (Fig.~2b -- it
is unresolved with a resolution of $43\arcsec$); its 
J2000 coordinates are:
$\alpha =4^{\rm h}58^{\rm m}30\fs1$,
$\delta=+47{\degr}58{\arcmin}34\arcsec$. Sh~217 is a low density 
\HII\ region, with an electron density of $\sim$50~cm$^{-3}$ deduced 
from the [O\,{\sc{ii}}] lines ratio, and of 20~cm$^{-3}$ from the 
radio emission. The shell morphology of Sh~217 may also be the 
signature of a champagne model. This morphology indicates that 
the electron density is higher farther from the star: in the 
champagne model the highest electron densities are found at the
borders of the cavity carved out of the high-density neutral material. 
However, in Sh~217, the velocities of the ionized gas and of the
neutral material are similar (Table~1); this suggests that, if an 
ionized flow exists, there is a large angle between the mean 
direction of flow and the line of sight.

\begin{figure*}
\centering
    \includegraphics[width=180mm]{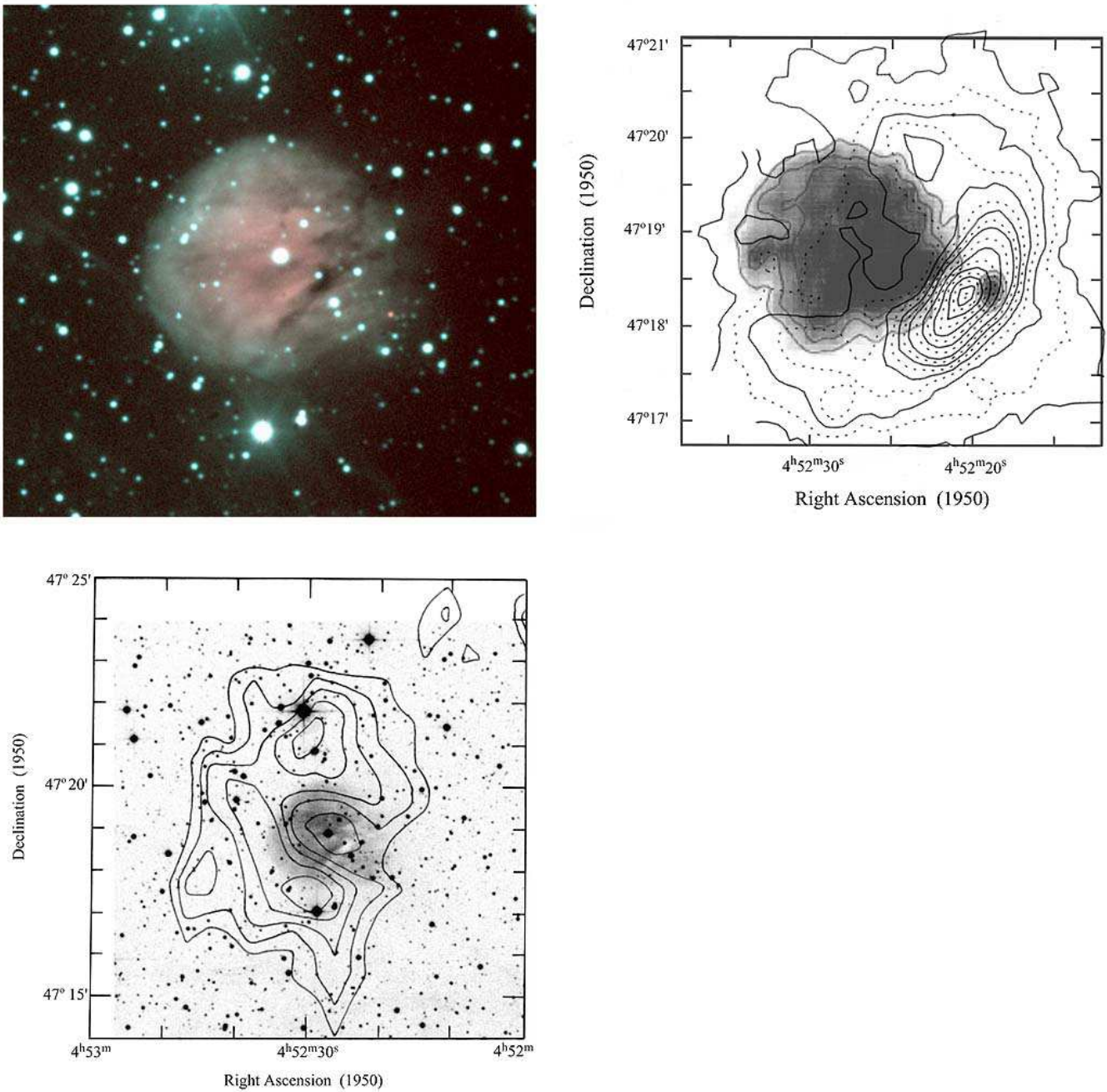}
\caption{
{\bf a)} Composite colour image of Sh~219 in the optical; pink
corresponds to the H$\alpha$ 6563~\AA\ emission, and turquoise to the 
[S\,{\sc{ii}}] 6717--6731~\AA\ emission (the [\SII] emission comes from
low-excitation ionized zones; it is enhanced close to the ionization
fronts). These frames were obtained with the 120~cm telescope of the
Observatoire de Haute-Provence (exposure times 1~h and 2~h
respectively). The size of the field is $5\farcm6 \times 5\farcm2$.
North is up and east is to the left.
{\bf b)} Sh~219 in the IR and radio continuum. The black contours 
(solid and dotted), drawn from the MSX 8.3~$\mu$m image, are
superimposed on the grey-scale 1.42~GHz continuum map of Leahy (1997). The MSX
radiance contours range from 1 to 14 $\times 10^{-6}$ W m$^{-2}$ sr$^{-1}$.
Note the half-ring of dust emission surrounding the \HII\ region.
{\bf c)} Integrated \HI\ emission associated with Sh~219 covering the
LSR radial velocity range from $-21.7$~km~s$^{-1}$ to $-26.6$~km~s$^{-1}$ 
(from Roger \& Leahy 1993). The \HI\ contours are superimposed
on the DSS2-red frame of Sh~219
}

\end{figure*}

\begin{figure*}
\centering
    \includegraphics[width=180mm]{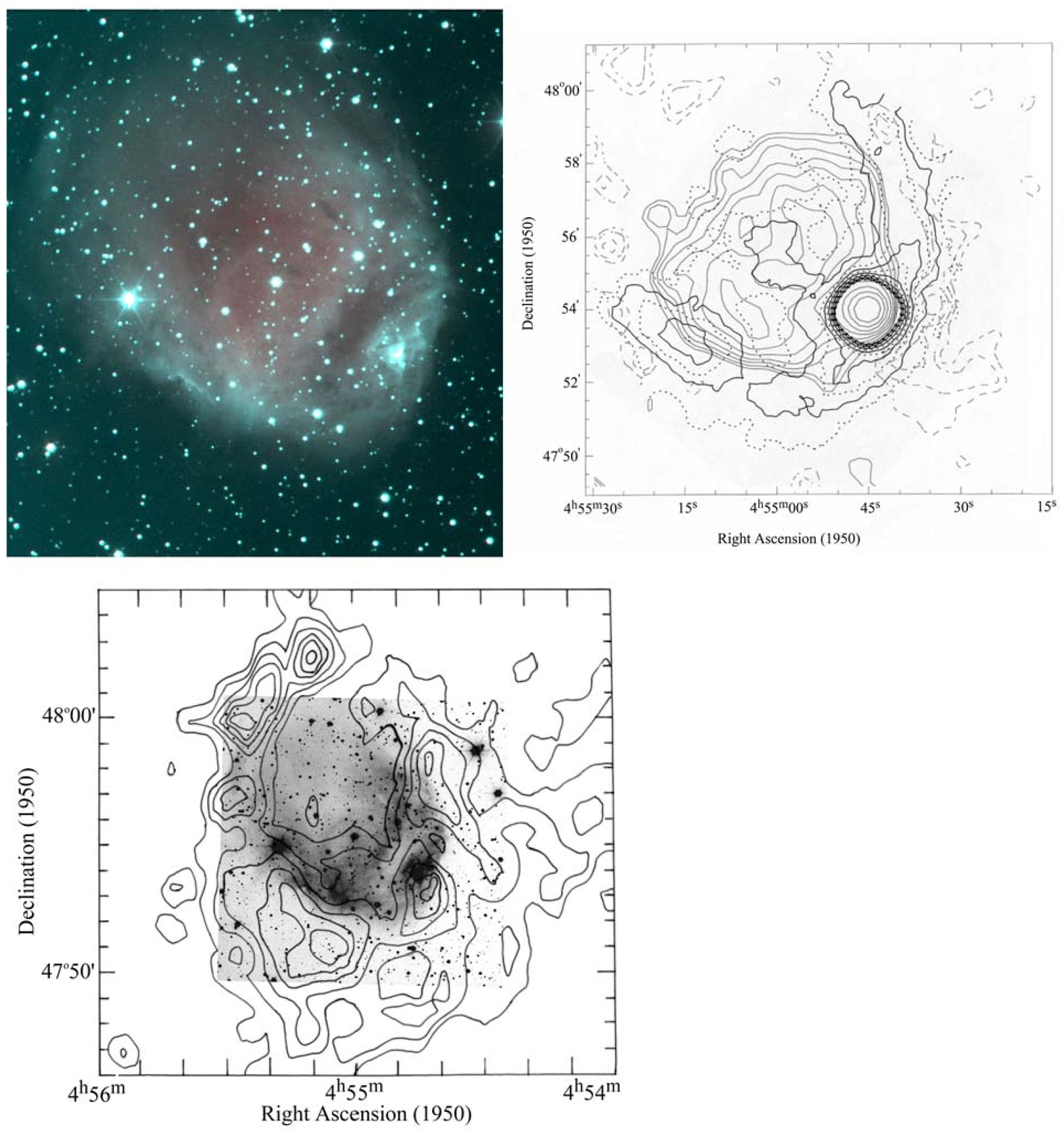}
\caption{
{\bf a)} Composite colour image of Sh~217 in the optical; same
characteristics as Fig.~1a. The size of the field is
$10\farcm25 \times 11\farcm25$.
{\bf b)} The black contours (solid and dotted), drawn from the MSX 
8.3~$\mu$m image, are superimposed on the radio continuum map of Sh~217 
(grey contours); the radio map is from the NVSS survey at 1.4~GHz (Condon
et al.\ 1998). Only those MSX radiance contours in the range from 0.6 to
3.0 $\times 10^{-6}$ W m$^{-2}$ sr$^{-1}$ are drawn; the radiance peaks at 
$30 \times 10^{-6}$ W m$^{-2}$ sr$^{-1}$ in the direction of the UC \HII\ region.
Note the half-ring of PAH emission surrounding the \HII\ region and the
good alignment, along the line of sight, of the radio and MSX
ultracompact sources.
{\bf c)} Integrated \HI\ emission associated with Sh~217 covering the
LSR radial velocity range from $-15.1$~km~s$^{-1}$ to $-20.1$~km~s$^{-1}$. 
The \HI\ emission is from Roger \& Leahy (1993). The \HI\
contours are superimposed on the DSS2-red frame of Sh~217
}
\end{figure*}

\subsection{The exciting stars}

Absolute integrated line fluxes of Sh~217 and Sh~219 in a number of
nebular emission lines were measured through a circular diaphragm of
diameter $4\farcm 5$ by Caplan et al.\ (2000). The He$^{+}$/H$^{+}$
ratios obtained for these \HII\ regions are rather low~-- 0.061 and
0.041 respectively~-- indicating that not all the helium is ionized;
the O$^{++}$ emission is very faint, with O$^{++}$/O ratios of
$5\times 10^{-2}$ and $6\times 10^{-3}$ respectively (Deharveng et
al.\ 2000). Thus, as regards their He$^{+}$ and O$^{++}$ emission, 
Sh~217 and Sh~219 appear to be low-excitation regions, very similar 
to Sh~153, Sh~168 and Sh~211, which are ionized by O9.5V, O9V and O9V
stars respectively. Also, Sh~219 is of lower excitation than Sh~217.
All this is consistent with a B0V spectral type for the main exciting
star of Sh~219, and suggests a slightly earlier spectral type,
possibly O9.5V or O9V, for the exciting star of Sh~217.

Sh~217 and Sh~219 are thermal radio continuum sources. Their flux
densities, measured at 1.46 GHz (Fich 1993, Roger \& Leahy 1993), are
in the ranges 400--472 mJy and 130--160~mJy respectively. We have
used eq.~1 of Simpson \& Rubin (1990) to determine the number of
Lyman continuum photons emitted per second by their exciting stars.
Assuming a distance of 5~kpc, helium ionic abundances of 0.06 for 
Sh~217 and 0.04 for Sh~219 and an electron temperature of 9200~K 
(Deharveng et al.\ 2000), we obtain $\log N_{\rm Lyc} = $ 48.0 for 
Sh~217 and 47.5 for Sh~219. According to Schaerer \& de Koter (1997)
these Lyman continuum fluxes indicate exciting stars of spectral
types B0V for Sh~217 and cooler than B0.5V for Sh~219. This is
somewhat cooler than the spectral types discussed in Sect.~2.1;
perhaps not all emitted Lyman continuum photons are detected, as is
expected in the case of a champagne model where the emission of the
low-density extended ionized component would probably not be
detected. However, the effective temperature and thus the Lyman
continuum flux of massive stars are still subject to revision 
(Martins et al.~2002), so that a difference of one subclass is
probably not significant.

\subsection{The photodissociation regions around Sh~217 and Sh~219}

Sh~217 and Sh~219 have been mapped in \HI\ at 21 cm, with an angular
resolution of $1\farcm 0 \times 1\farcm 4$ (Roger \& Leahy 1993). 
Both \HII\ regions show prominent photodissociation regions (PDRs)
surrounding the ionized zones.

Figure~2c shows the \HI\ emission associated with Sh~217, integrated
from $-15.1$ to $-20.1$~km~s$^{-1}$ ($V_{\rm LSR}$), superimposed on a
DSS2-red frame of this region. An almost complete ring of \HI\
emission surrounds the \HII\ region. The ring is broad, with a
typical radial extent of 5--10~pc and with a peak column density of 3
$\times 10^{20}$~cm$^{-2}$. The ring contains about 650~$M_{\odot}$
of atomic hydrogen, with a mean density of 15~atoms~cm$^{-3}$.

Figure~1c shows the \HI\ emission associated with Sh~219, integrated
from $-21.7$ to $-26.6$~km~s$^{-1}$, superimposed on a DSS2-red frame.
The surrounding \HI\ emission is less symmetric than for Sh~217, with
a pronounced deficiency on the western side, in the direction of the
stellar cluster discussed in Sect.~3.2. This incomplete ring has a
peak column density of 1.8$\times 10^{20}$~cm$^{-2}$, and contains
about 86~$M_{\odot}$ of atomic hydrogen with a mean density of 
9~atoms~cm$^{-3}$.

We have searched for Sh~217 and Sh~219 in the MSX infrared survey 
(Egan et al.\ 1999; this survey has an angular resolution of 
18$\arcsec$ in band A, centred at 8.3 $\mu$m). These \HII\
regions have counterparts in MSX's band A,
where they appear as semi-circular emission rings (Figs~1b and
2b). In each case an `MSX point source' is observed in the
direction of the ring. Their positions and fluxes are given in 
Table~2.

The MSX band A covers the 6.8--10.8~$\mu$m range (at half
transmission). It contains the 7.7~$\mu$m and 8.6~$\mu$m emission
bands, often attributed to polycyclic aromatic hydrocarbons (PAHs;
Léger \& Puget 1984, Alamandola et al.\ 1985; see also Verstraete et
al.\ 2001 for a discussion of the origin of these bands), and an
underlying continuum attributed to very small grains. 
PAHs cannot survive in ionized regions. This is
clearly demonstrated by the ISO spectra obtained at different
positions in the M~17 field by Cesarsky et al.\ (1996; their fig.~2);
the spectrum obtained in the direction of the ionized region is
dominated by a continuum, rising strongly with increasing wavelength,
whereas that obtained in the direction of the neutral PDR shows the
PAH emission bands superimposed on a continuum. In the PDR, the PAHs
absorb the strong UV radiation leaking from the \HII\ region, are
heated (possibly ionized), and radiate in the PAH bands. This is most
probably the origin of the band A emission forming
the rings surrounding the Sh~217 and Sh~219 \HII\ regions.

The MSX point sources coincide with the ultracompact
radio sources
and the infrared clusters discussed in this paper. These `point
sources' are resolved: that associated with Sh~217 has circular
isophotes and, after correction for the PSF, a FWHM of 25$\arcsec$ 
(0.6~pc); that associated with Sh~219 is elongated along the
ionization front, with a FWHM perpendicular to the front of 
$11\arcsec$ (0.25~pc). These MSX point sources are also IRAS point
sources. Table~2 gives their IRAS identifications, fluxes and
luminosities (calculated according to Chan \& Fich 1995). IRAS~04547+
4753, associated with Sh~217, has the characteristic colours of an ultracompact
\HII\ region (Wood \& Churchwell 1989). Its low-resolution IRAS
spectrum shows a red continuum with no PAH features (Chen et al.\
1995). Its luminosity
is that of a cluster containing a few early B stars. The colours and
luminosity of IRAS~04523+4718, associated with Sh~219, are more
characteristic of intermediate mass protostars (e.g.\ Herbig Be
stars) than of ultracompact \HII\ regions. No maser emission has been detected
in the direction of these two IRAS sources (Wouterloot et al.\ 1993,
MacLeod et al.\ 1998, Szymczak et al.\ 2000).

\begin{table*}
\caption{MSX and IRAS observations of the point sources associated with Sh~217
 and Sh~219}
\vspace{3mm}
\begin{tabular}{cccc@{\hspace{2mm}}c@{\hspace{2mm}}c@{\hspace{2mm}}c@{\hspace{2mm}}c@{\hspace{2mm}}c}
 \hline\hline
       &        &        & \multicolumn{5}{c}{Flux (Jy)} & \\
 Source & $\alpha$(2000) & $\delta$(2000) & 8.3 $\mu$m  & 12 $\mu$m & 
 25 $\mu$m & 60 $\mu$m & 100 $\mu$m & $L_{\rm IR}$ ($L_{\odot}$) \\
 \hline
MSX Sh~217 & $4^{\rm h}58^{\rm m}30\fs3$ & $+47{\degr}58{\arcmin}33\arcsec$ & 11.6 & & & & & \\
IRAS04547+4753 & $4^{\rm h}58^{\rm m}29\fs7$ & $+47{\degr}58{\arcmin}28\arcsec$ & & 10.42 & 82.03 & 359.80 & 367.30 & 22700\\ 
 \hline
MSX Sh~219 & $4^{\rm h}56^{\rm m}03\fs9$ & $+47{\degr}22{\arcmin}55\arcsec$ & 1.25: & & & & & \\
IRAS04523+4718 & $4^{\rm h}56^{\rm m}04\fs0$ & $+47{\degr}22{\arcmin}58\arcsec$ & & 3.30 & 5.21 & 81.70 & 304.70 & 5900 \\
 \hline
\end{tabular}
\end{table*}

\subsection{The molecular material associated with Sh~217 and Sh~219}

Sh~217 has been mapped in $^{12}$CO and $^{13}$CO (1--0 transition),
with a resolution of 65$\arcsec$, by Jackson \& Sewall (1982). The
4100~$M_{\odot}$ of molecular material associated with Sh~217
is located mainly in three components. The main component is
associated with the ultracompact \HII\ region and the MSX point source. In this
direction, the mean molecular hydrogen density is 730~cm$^{-3}$, and
the column density $N$($^{13}$CO) reaches $9 \times 10^{15}$~cm$^{-2}$
(corresponding to $N($H$_2) \sim 4.5 \times 10^{21}$~cm$^{-2}$ according
to Jackson \& Sewell). The low angular resolution does not allow us
to see any details connected with the ionization front or with the
cluster. The visual extinction corresponding to this column density,
$A_V\sim 4.5$~mag (Thronson et al.~1985), is much smaller than that
observed for individual stars in the cluster (Sect.~3), suggesting
that the dust/gas ratio is anomalous or, more probably, that dense
structures are present but not observed. Two fainter components are
observed in the direction of the \HII\ region (Jackson \& Sewall
1982). Comparison of the \HI\ and CO emission maps shows a
remarkable anti-correlation between these two emissions.
 
Very little molecular material is found to be associated with Sh~219 
(Huang \& Thaddeus 1986).

\begin{figure*}
\centering
  \includegraphics[width=180mm]{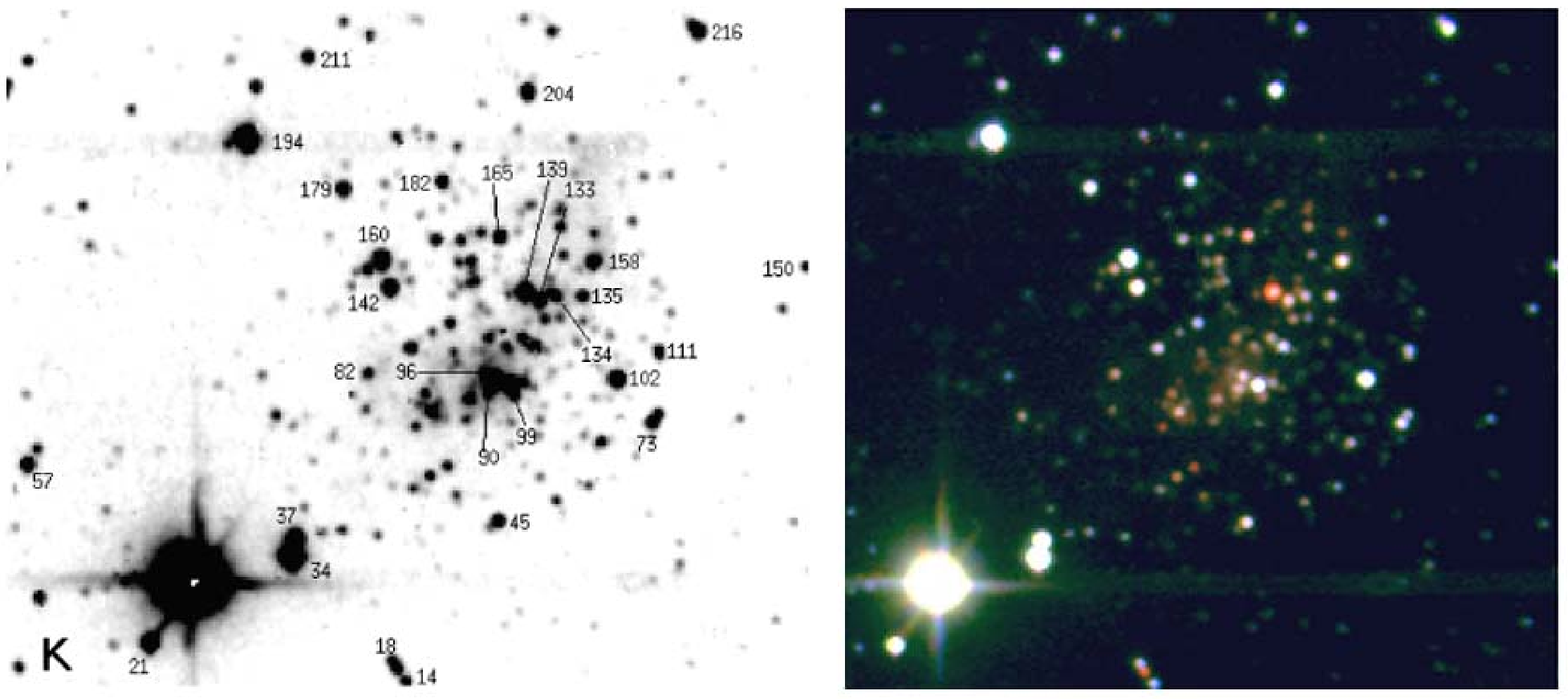} 
 \caption{{\bf a)} $K$ frame of the Sh~219 field. The frame size is
$3\farcm 35 \times 2\farcm 8$. North is up, east is to the left. The stars
discussed in the text are identified by their numbers in Table~3.
{\bf b)} $JHK$ colour composite of the cluster associated with Sh~219
(blue for $J$, green for $H$ and red for $K$)}
\end{figure*}

\begin{figure*}
    \includegraphics[width=180mm]{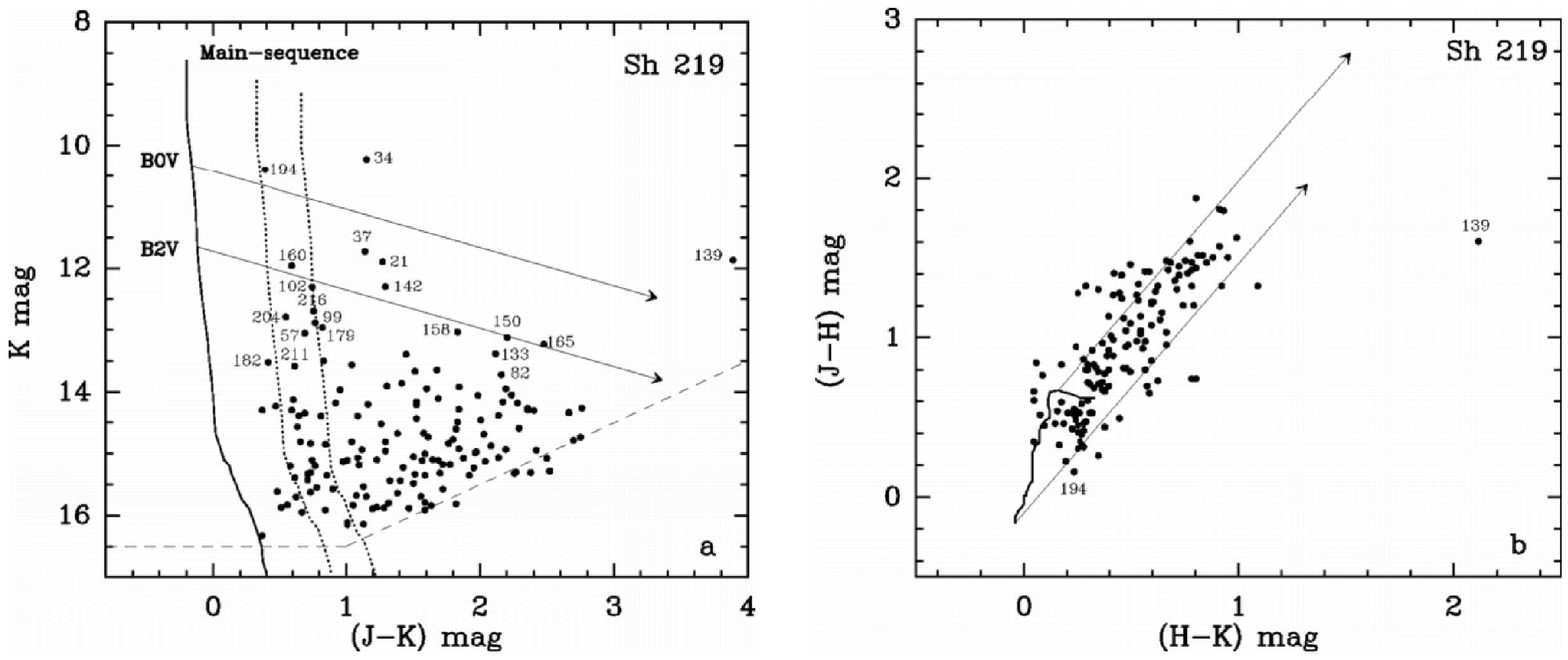} 
\caption{
{\bf a)} The $K$ versus $J-K$ magnitude-colour diagram of stars in
the Sh~219 field. The main sequence is from Vacca et al.\ (1996) for
O stars and from Schmidt-Kaler (1982) for later spectral types. The
main sequence is drawn for a distance of 5~kpc, and for visual
extinctions of zero (solid curve), 3 and 5 mag (dotted curves).
The dashed lines show our detection limits. Reddening lines are drawn
originating from B0V and B2V stars; they correspond to a visual
extinction of 20 mag and the standard extinction law of Mathis
(1990) with $R_{V}=3.1$.
{\bf b)} The $(J-H)$ versus $(H-K)$ colour-colour diagram. The main
sequence and the reddening lines corresponding to a visual extinction
of 20 mag are shown as in Fig.~4a
}
\end{figure*}

\section{Star formation associated with Sh~217 and Sh~219}

\subsection{Near-IR observations}

Frames of these regions in $J$, $H$ and $K'$ have been obtained
in October 2000 at the
2.12-m telescope of the Observatorio Astronómico Nacional at San Pedro
Mártir, Baja California, México. We have used the CAMILA camera 
(Cruz-González et al.\ 1994) with the f/4.5 focal reducer, which
gives a scale of $0\farcs85$ per pixel and a field of
$3\farcm6 \times 3\farcm6$. The observing and reduction techniques are
as described in Deharveng et al.\ (1997).

Eighteen UKIRT standard stars, covering a large colour range 
($-0.231 \leq J-K \leq +2.992$), were observed to determine the colour 
equations and zero points of the photometric system. In the following, 
all the magnitudes are given after transformation to the UKIRT system.

\subsection{Results for Sh~219}

Sh~219 was observed in October 2000 with total integration times of
750~s in each band. Our magnitude detection limits are of about 17.5 
in $J$ and 16.5 in $H$ and $K$ (corresponding to a threshold of 
$3 \sigma$; 222 stars were measured in at least two bands, of
which 142 were measured in all three bands. The J2000 coordinates
of these stars, their $K$ magnitudes and their $J-K$, $J-H$ and $H-K$
colours are given in Table~3. The uncertainty of each magnitude is 
generally smaller than 0.2~mag.

\begin{table}
\caption{Coordinates and $JHK$ photometry of all the stars in a
3\farcm5 $\times$ 3\farcm5 field centred on Sh~219. This Table, 
available at the CDS, contains the following information. Column 1 gives the 
numbers of the stars, Columns 2 and 3 their $X$ and $Y$ tangential coordinates in
arcminutes, Columns 4 through 9 their J2000 equatorial coordinates, Column 10 
the $K$ magnitude and Columns 11 through 13 the colours $J-K$, $J-H$ and $H-K$.}
\end{table}

Figure~3a shows the $K$ frame. The stars discussed in the text are
identified by the same numbers as in Table~3. Figure~3b is a $JHK$
colour composite image. This image shows the effects of extinction,
the differences in colour of the stars mainly being due to
differences in reddening. These figures show two important facts: 
i)~a previously unreported cluster of reddened stars appears at the 
south-west periphery of the Sh~219 \HII\ region; ii)~the central 
exciting star of Sh~219, no.~194, appears
isolated: no nearby red companions appear in $K$.

Figures~4a and 4b present respectively the $K$ versus $J-K$ and the
$J-H$ versus $H-K$ diagrams. Figure~4b shows
that most of the stars appear to be reddened main-sequence stars,
and thus Fig.~4a can be used to estimate their visual extinctions.
Certain stars -- nos~99, 102, 160, 179, 194, 204, 216 -- are
affected by a relatively small visual extinction (3--5 mag). This
probably corresponds to the general interstellar extinction for the
distance of Sh~219; these stars are probably unaffected by local
extinction. The exciting star of Sh~219, no.~194, whose near-IR 
magnitudes are consistent with a B0V spectral type
(see Sect.~2.1), belongs to this
group. The case of stars 21, 34, 37 and 142 is not as
clear. If these are main-sequence stars, they have a visual
extinction in the 7.2--8.1 mag range; if situated at the distance of
Sh~219, they are intrinsically luminous, and star no.~34 at least
should ionize the surrounding medium and form an \HII\ region. Such an
\HII\ region is detected neither in the DSS2-red survey nor in the
radio continuum maps. Perhaps these stars are cool giants unrelated to the
cluster; their positions in the $J-H$ versus $H-K$ diagram are
compatible with this possibility. A number of stars -- nos~82, 133, 139, 150, 158,
165, etc.\ -- mostly located in the direction of the small cluster,
have extinctions greater than 11 mag. Star 139, close to
the centre of the cluster, is one of the brightest and most highly
reddened objects, with $A_V \sim 24$ mag. It is intrinsically
brighter in $K$ than no.~194, the exciting star of Sh~219. Star~139
displays a clear near-IR excess; it is probably still
associated with very hot dust situated in an accreting disk or
envelope, very close to the central object. The H$\alpha$ +
[S\, {\sc{ii}}] colour image of Sh~219 (Fig.~1a) shows that 
star 139 is
also an H$\alpha$ emission star (it is the only `pink star' of the
image). Star no.~139 cannot be the exciting star of the non-resolved
radio continuum source, as it lies 18$\arcsec$ away. Star 158, lying
1$\arcsec$ from the radio source (visual extinction 
$\sim$11.5~mag), is a possibility.

The cluster  observed at the south-west periphery of Sh~219 subtends
an angle of about 60$\degr$ with respect to the exciting star of 
Sh~219. The cluster is not spherical, but is elongated along the
ionization front, suggesting that star formation was triggered by the
passage of the front (see Sect.~4.4).

\subsection{Results for Sh~217}

The stellar cluster discovered at the periphery of Sh~217 has been studied in the 
near IR as part of a large survey of the Initial Mass Function in 
young clusters by Porras (2001) and Porras et al.\ (2000, 2002). $JHK$ photometry 
and coordinates of 121 stars are presented here. Their 
coordinates and magnitudes are given in Table~4. 
The stars discussed in the text are identified on the $H$ frame, in Fig.~5a. 

\begin{figure*}
\centering
    \includegraphics[width=180mm]{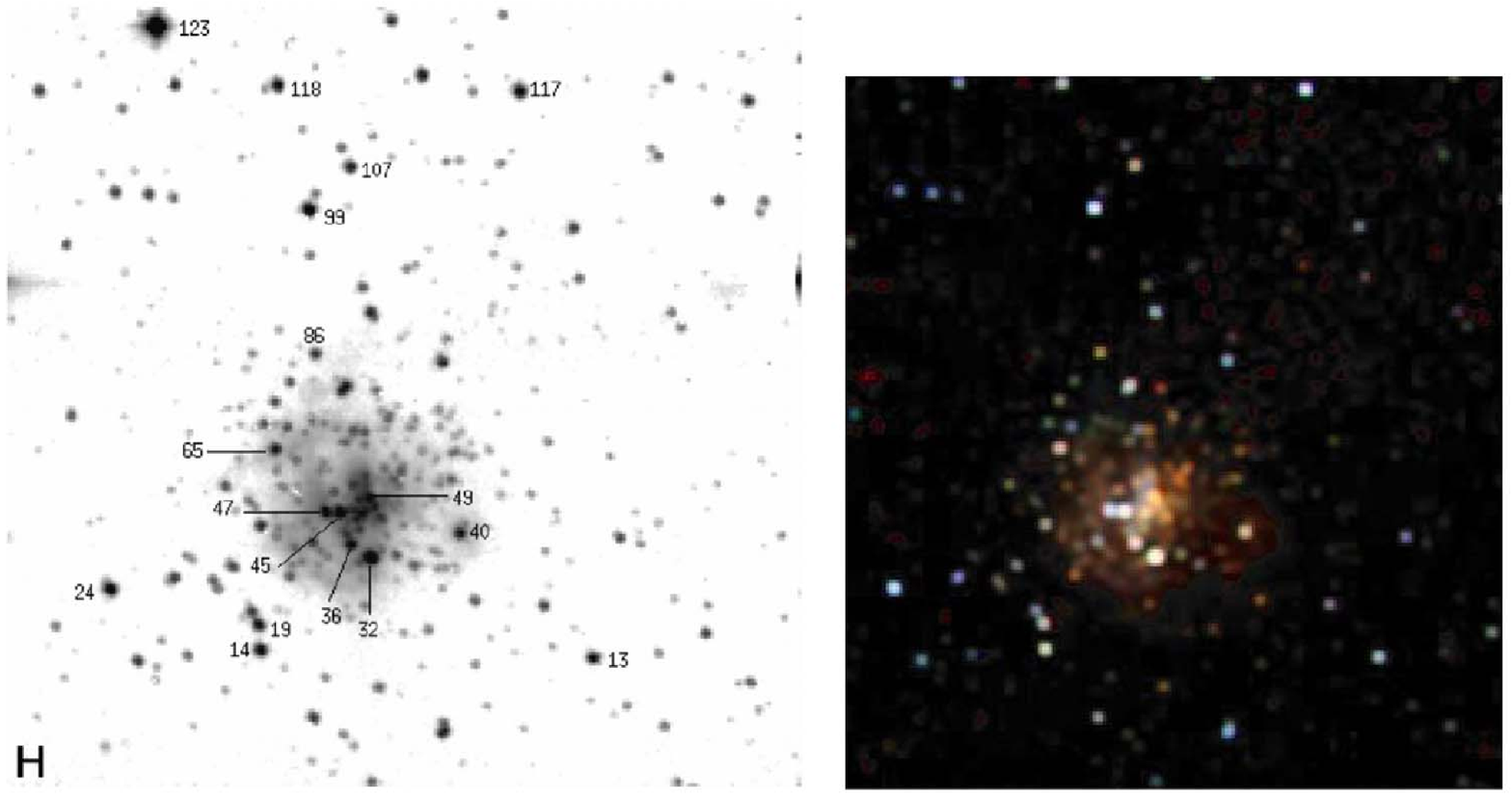} 
\caption{
{\bf a)} $H$ frame of the Sh~217 field. The field size is
$3\farcm 3\times 3\farcm 4$. North is up, east is to the left. The stars
discussed in the text are identified by their numbers from 
Table~4.
{\bf b)} Composite colour image of the young stellar cluster observed at 
the periphery of Sh~217. 
$J$ is blue, $H$ is green and $K$ is red\vspace*{15mm}
}
\end{figure*}

Figure~5a shows an almost spherical cluster, with luminous objects at
its centre; 90\% of the stars in the cluster are enclosed in a circle
of radius 1.1~pc. Figure~5b is a $JHK$ colour composite image of this
cluster and the surrounding stars. The image emphasizes the large
variations of extinction among the stars in this field, especially among the
massive objects in the direction of the centre of the cluster. 
region.

The colour-magnitude diagram is presented in Fig.~6. The bright star 123, with 
$A_V$=0.9~mag, is probably a foreground star. Stars 24, 47, 99 and 117 have
$A_V$ in the range 2--4~mag, whereas stars 13, 14, 19, 36, 40, 45, 65,
107 and 118 are more reddened, with $A_V$ from 5 to 11~mag. The two blue stars, 
nos~45 and 47 ($A_V$$\sim$ 
5.6~mag and 3.4~mag respectively), have been tentatively classified as B2V by Moffat et al.\ 
(1979) on the basis of their $UBV$ photometry. The colour-magnitude 
diagram confirms 
that they are B stars, but probably of a later spectral type, in which case they 
are not the only ionization sources of the associated compact \HII\ 

\begin{figure}
\centering
    \includegraphics[width=88mm]{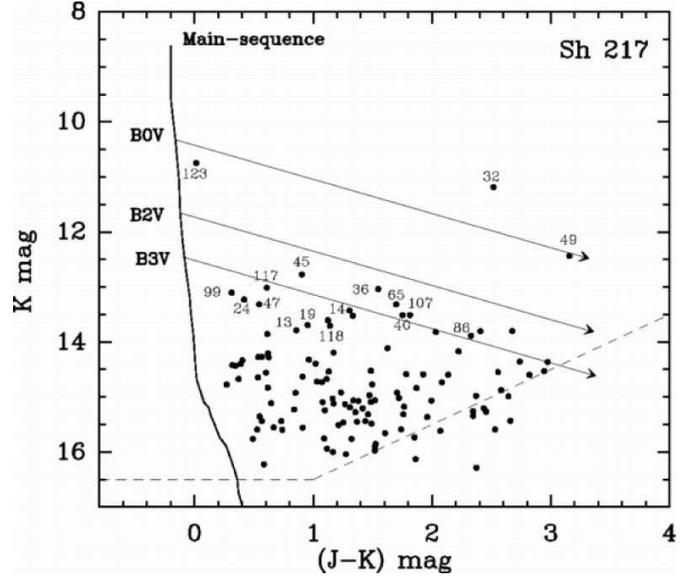} 
\caption{The $K$ versus $(J-K)$ magnitude-colour diagram of stars in
the Sh~217 field. Main sequence, detection limits and reddening as in
Fig.~4a}
\end{figure}

The mean visual extinction of the cluster is about 8~mag. The western
part of the cluster shows deeply embedded objects. Two red objects,
nos~32 and 49, appear very luminous in $K$, as bright as B0V stars
(Fig.~6).
Assuming they are on the main sequence, their visual extinctions
would be 15.5~mag and 19.1~mag respectively. However star 32 shows a
near-IR excess, and thus is probably a Herbig Be star with less
extinction. Star 49 presents no near-IR excess, but its $J$
magnitude is uncertain (due to the high stellar density, the high
background emission and the high extinction, the uncertainty in the
colours of the faint objects in this cluster may be greater than 
0.3~mag). These two stars probably contribute to the ionization of the
associated compact \HII\ region.

\begin{table}
\caption{Coordinates and $JHK$ photometry of all the stars in a 
3\farcm1 $\times$ 3\farcm2 field centred on Sh~217. This Table, 
available at the CDS, contains the following information. Column 1 gives the 
numbers of the stars, Columns 2 and 3 their $X$ and $Y$ tangential coordinates in
arcminutes, Columns 4 through 9 their J2000 equatorial coordinates, Column 10 
the $K$ magnitude and Columns 11 through 13 the colours $J-K$, $J-H$ and $H-K$.}
\end{table}

\section{Discussion}

\subsection{How old are Sh~217 and Sh~219?}

The dynamical ages of Sh~217 and Sh~219 can be estimated by using the 
model by Dyson \& Williams (1980) of an \HII\ region expanding in a 
homogeneous medium. An unknown parameter
here is the density of the medium into which the \HII\ regions
evolve. Sh~219 is excited by a B0V star, which emits
$1.25 \times 10^{48}$ ionizing photons per second (Vacca et al.\
1996, Schaerer \& de Koter 1997). If expanding into a homogeneous
medium of $10^3$~cm$^{-3}$, Sh~219's present radius of 2.2~pc would
correspond to an age of 4.4~$\times$10$^5$ years and an expansion
velocity of 2.7~km~s$^{-1}$. If expanding into a lower density
medium, for example $10^2$~cm$^{-3}$, it would be much younger -- 
6.6~$\times$10$^4$ years -- and would be expanding with the higher
velocity of 8.5~km~s$^{-1}$.

Sh~217 is twice as large as Sh~219. Given that it is excited by a
star emitting $2.0 \times 10^{48}$ ionizing photons per second, its
larger size implies that it is either older than Sh~219 or that it
has developed in a lower density medium. For example, expanding into
a homogeneous medium of $10^3$~cm$^{-3}$, its dynamical age would be
1.4~$\times$10$^6$ years and its present expansion velocity 
1.8~km~s$^{-1}$. Alternatively, an age of 4.4~$\times$10$^5$ years 
implies expansion in a $150$~cm$^{-3}$ medium, with a present expansion
velocity of 4.6~km~s$^{-1}$.

Most probably the exciting stars of these regions formed in high density
molecular cores but, later, the associated \HII\ regions evolved within a
lower density medium. Furthermore, all of these estimations assume that
both the \HII\ region and its surrounding medium are spherically
symmetric around the exciting star. As we have seen in Sect.~2.2, there
is some kinematic evidence that Sh~219 is undergoing a champagne flow,
and the shell morphology of Sh~217 may also be the signature of a
champagne model. If these \HII\ regions correspond to a champagne model,
they are older than estimated previously, and their expansion velocities
lower. This would explain why no expansion is observed in the \HI\ rings
surrounding these \HII\ regions (Roger \& Leahy 1993).

The relative sizes and masses of the \HII\ regions and of their atomic PDRs
are also age indicators. They depend 
mainly on the exciting stars, on the ambient density, on the amount of
photoabsorption by dust grains, and are functions of the time. 
As a region expands,
the dissociation rate of molecular hydrogen behind the ionization front,
which depends on the UV energy density, decreases. As a consequence,
the dissociation front is eventually caught up by the ionization
front, and the \HI\ zone disappears (Rogers \& Dewdney 1992).

According to the observations, the ratios of the sizes and masses of the
photoionized and photodissociated regions of Sh~217 and Sh~219 are 
respectively 1.9 and 2.2 for 
$R_{\rm H\,{\scriptscriptstyle I}}/R_{\rm H\,{\scriptscriptstyle II}}$,
and 1.9 and 2.5 for 
$M_{\rm H\,{\scriptscriptstyle I}}/M_{\rm H\,{\scriptscriptstyle II}}$.
Comparison with the non-stationary models of Roger \& Dewdney 
(1992) indicates that
Sh~217 and Sh~219 cannot have developed in a medium as dense as 
$10^3$~cm$^{-3}$, but most probably in a medium of density 
$\leq 30$~cm$^{-3}$.
But then, the ages predicted by the models are incredibly
young (a few $10^4$ years; Roger \& Leahy 1993).

Dust absorption is important. This effect reduces the size
of the ionized and atomic zones, as well as the
$R_{\rm H\,{\scriptscriptstyle I}}/R_{\rm H\,{\scriptscriptstyle II}}$ and 
$M_{\rm H\,{\scriptscriptstyle I}}/M_{\rm H\,{\scriptscriptstyle II}}$
ratios; dust absorption is more important in the photodissociation
zone, as shown by the models of Diaz-Miller et al.\ (1998). 
Weingartner \& Draine (2001) have proposed that the stellar
radiation pressure drives the grains through the gas, leading to an 
increase of the dust-to-gas ratio in the PDR. Large amounts of dust
in the PDRs of Sh~217 and Sh~219 could account, at least partly,
for the low values of the observed ratios. This last effect, not
taken into account in the models, as well as the possibility of a clumpy PDR, 
result in very uncertain age determinations.

\subsection{The great similarity of Sh~217 and Sh~219}

Although Sh~217 and Sh~219 are separated in projection by some 65~pc,
they are similar in many respects: i) Sh~217 and Sh~219 appear as
spherical (or hemispherical) \HII\ regions, excited by a single central star
(see Sect.~2.3); ii) the \HII\ regions are surrounded by annular \HI\
zones of low column density (accounting for less than 0.5~mag of
visual extinction); iii) a cluster is observed at one position at the
periphery of each \HII\ region, in the direction of the PDR, in an
area of high dust emission very close to the ionization front; iv)
these clusters are reddened, with some objects affected by a visual
extinction up to 20~mag, and they contain massive stars, which
ionize compact \HII\ regions; v) the offset direction
of the cluster from the exciting star of the extended \HII\ region is
the same, to within a few degrees, for Sh~217 and Sh~219: these
directions are nearly perpendicular to the Galactic plane. 
Fig.~7 summarises this situation.

\begin{figure}
\centering
    \includegraphics[width=88mm]{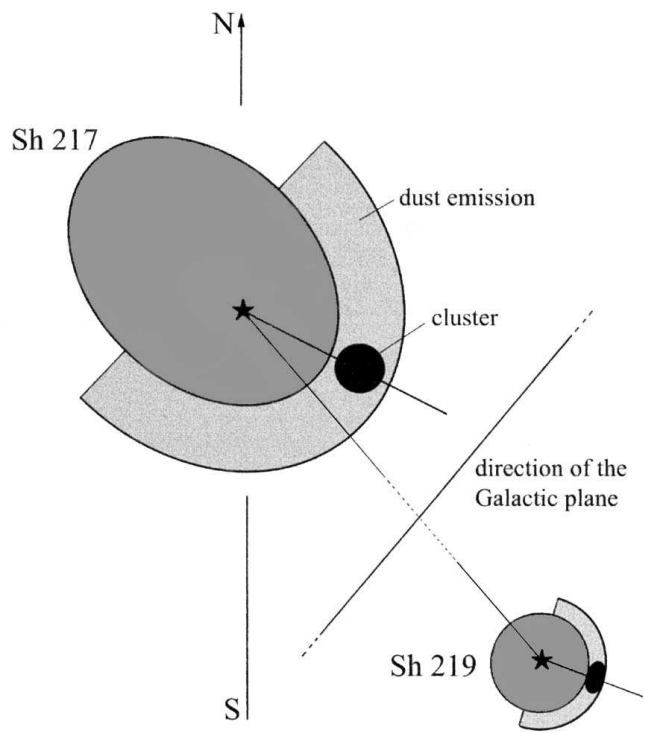} 
\caption{A schematic view of the respective positions and dimensions of the 
Sh~217 and Sh~219 \HII\ regions, their half-rings of PAHs emission, and the 
associated embedded clusters. The distance between Sh~217 and Sh~219 is not 
to scale}
\end{figure}

A difference between the two \HII\ regions is that very 
little molecular material seems to be associated with Sh~219, contrary 
to the situation in Sh~217.  

\subsection{Sequential star formation?}

As far as we can see from the present observations, the exciting
stars of Sh~217 and Sh~219 are isolated massive stars: no nearby
luminous companions appear on the $K$ frames of these regions as they
do, for example, for the exciting stars of Sh~138 (Deharveng et al.\
1999). Have these stars formed in isolation?

If we assume that the exciting stars of Sh~217 and Sh~219 formed in a
common cluster and were both subsequently ejected after internal
dynamical interactions $t$($10^6$)~years ago, their current
separation (65~pc in projection) implies space velocities larger than
{\mbox {$32/t$~km~s$^{-1}$}}. This value seems somewhat high for
massive stars. Also, if the stellar velocity dispersion is 
2~km~s$^{-1}$ (as in the Orion cluster, Jones \& Walker 1988), then 
$t$~million~years after its formation this cluster should have a
diameter of $2.1 \times t$~pc. We have examined the DSS and
2MASS surveys, trying to detect this cluster by eye from an over-density
of stars at some location between Sh~217 and Sh~219, without
success.

Another possibility is to assume that the exciting stars of Sh~217
and Sh~219 formed in their respective peripheral clusters which we
observe today, and were subsequently ejected from these clusters. There
is a finite probability that the exciting 
star of each \HII\ region formed in the
cluster and was ejected; but the probability that, being
ejected, each forms an \HII\ region with a radius just equal to the
star-cluster distance is very low. This requires that, for each
region, the ejection velocity be, by chance, equal to the radius of
the \HII\ region divided by its age; furthermore, as the two \HII\
regions present the same configuration (same position angle
$\pm 10\degr$ of the cluster with respect to the \HII\ region's
exciting star), the direction of the ejection must be, by chance, the
same for both regions!

Thus the clusters observed at the periphery of the Sh~217 and Sh~219 
\HII\ regions are most probably second-generation clusters.

\subsection{Physical processes of sequential star formation}

In the following we assume that we are observing second-generation
clusters, situated at the present time inside the PDRs. The best way
to test this idea would be to compare the age of the \HII\ regions
and that of the clusters but, as we are dealing with young massive
objects evolving rapidly, it is almost impossible to estimate ages
accurately enough to allow comparison.

According to the model of Elmegreen \& Lada (1977; cf.\ also
Whitworth et al.\ 1994), as the \HII\ region expands, a thin layer of
compressed neutral material forms between the ionization front and
the shock front that precedes it in the neutral material. This layer
may eventually become gravitationally unstable, fragment and form
stars. This may be the configuration observed in Sh~217 and Sh~219.
The elongation along the ionization front of the cluster associated
with Sh~219 is a strong argument in favour of this model.  

The \HI\ zones surrounding Sh~217 and Sh~219 do not correspond to
such a compressed layer. Whitworth et al.\ (1994) have shown that the
layer fragments when its column density reaches $\sim$6~10$^{21}$ cm$^{-2}$. 
This is higher than the observed column density in the \HI\
rings by a factor $\sim$20. Thus if such a compressed layer exists it
is most probably molecular. High angular resolution observations of
the molecular material associated with Sh~217 and Sh~219 would
confirm or infirm the existence of this layer. If this model were
confirmed, we would need to explain how a wide low-density \HI\ zone
as well as a thin layer of dense molecular material can surround 
these \HII\ regions. Can a clumpy PDR solve this problem (with dense
molecular clumps surrounded by low-density atomic gas)?

A problem with Elmegreen \& Lada's model concerns the ages and sizes
of the \HII\ regions. Using the same numbers as in Sect.~4.1 to
estimate the ages of Sh~217 and Sh~219, we find that these \HII\
regions (especially Sh~219) are too young for the layer to become
gravitationally unstable and form stars. If the \HII\ regions satisfy
the champagne model, they are older; can this particular geometry
help solve this problem? Another possibility is that the 
shocked expanding layer, prior
to the beginning of the instability, collides with a pre-existing
molecular core. Star formation would take place at the interface
between the layer and the cloud core. This would explain why only one
site of star formation is observed along each ring, and could explain
why the young cluster of Sh~217 is observed at the border of both the
\HII\ region and the main molecular core.

Whitworth et al.\ (1994) have shown that, when the fragmentation of
the shocked layer starts at the border of the expanding \HII\ region,
the mass of the fragments 1) depends very little on the ionizing
flux, 2) varies as $n^{- \frac {5}{11}}$, where $n$ is the density of
the surrounding medium and 3) depends strongly on the turbulence in
the shocked layer, and thus on the temperature $T$ of the gas in this
layer, varying as $T^{\frac {20}{11}}$. PDR models (see Hollenbach \&
Tielens~1997 and references therein) predict high temperatures for
the gas, as high as 1000~K close to the ionization fronts. This
situation favours the formation of massive fragments susceptible to
form massive stars and possibly clusters.

Sh~217 and Sh~219 are not the only such objects. We have used the MSX
survey to search for annular (or semi-annular) dust emission
structures surrounding small \HII\ regions having MSX point sources
in the direction of the dust rings. We have found about twenty such
objects (Deharveng et al., in preparation). Most of these MSX point
sources are also IRAS point sources, and some are associated with
embedded clusters. These sources are possibly second-generation
clusters.

\section{Conclusions}

Sh~217 and Sh~219 are textbook examples of the appearance of an \HII\
region excited by a central isolated massive star (a Strömgren
sphere) and its associated photodissociation region. We have argued
that the physical conditions present in these photodissociation
regions have favoured the formation of second-generation clusters
containing massive stars. The case of these two regions is not
unique.

However, as often, an apparently simple configuration turns out to be
more complicated than anticipated. Here the \HII\ regions are
possibly non-spherical, and are best described by a champagne model;
the \HI\ PDR is possibly inhomogeneous. We have
presented observational evidence suggesting that sequential star 
formation occurs at the periphery of Sh~217 and Sh~219. However, we are
still unable to demonstrate what physical process is at work --
whether gravitational instability of the shocked layer surrounding
the \HII\ regions or collision of this expanding layer with a 
pre-existing molecular core. We also need to determine which
physical conditions allow the formation of clusters: high temperature
and/or inhomogeneity of the PDR, slow evolution of non-spherical \HII\ 
regions, etc. High resolution molecular observations
of the PDRs of these regions should allow us to answer some 
of these questions.

\begin{acknowledgements}

We would like to thank the San Pedro Mártir Observatory staff for
their support during the observations. D.~Gravallon is thanked for
the optical frames of Sh~217 and Sh~219 that he obtained for this
study at the Observatoire de Haute-Provence. The MSX Survey has been
particularly useful, and we thank M.~Egan and S.~Carey for their
help. The referee's useful comments are greatly appreciated.
This publication makes use of the NRAO VLA Sky Survey, and of
the NASA/IPAC Infrared Science Archive, which is operated by the Jet
Propulsion Laboratory, California Institute of Technology, under
contract with the National Aeronautics and Space Administration. This
research has made use of the Simbad astronomical database operated at
CDS, Strasbourg, France. L.S.\ and A.P.\ thank the Université de
Provence and the Laboratoire d'Astrophysique de Marseille
for providing one month of financial support in Marseille.

\end{acknowledgements}

\end{document}